\documentclass[epj]{svjour}

\usepackage{graphicx}
\usepackage{fancyhdr}
\def\makeheadbox{{
 }}
\def\mytitle{My title} 
\def\myauthors{My name}  
\def\mytype{My type of session}
\def\mysession{My session}

\def\mytitle{Precision Measurements of the Stop Quark Mass at the ILC~~~~~\thepage} 
\def\myauthors{Andr\'e Sopczak}
\def\mytype{\thepage}
\def\mysession{Colliders - SUSY Phenomenology}

\def\fb{\mathrm{\,fb}}

\def\st{\mbox{$\rm \tilde{t}_1$}}

\def\ifmath#1{\relax\ifmmode #1\else $#1$\fi}%
\def\chio{\ifmath{\mathchoice%
     {\displaystyle\raise.4ex\hbox{$\displaystyle\tilde\chi{^0_1}$}}%
        {\textstyle\raise.4ex\hbox{$\textstyle\tilde\chi{^0_1}$}}%
      {\scriptstyle\raise.3ex\hbox{$\scriptstyle\tilde\chi{^0_1}$}}%
{\scriptscriptstyle\raise.3ex\hbox{$\scriptscriptstyle\tilde\chi{^0_1}$}}}}

\newcommand{\est}{\epsilon_{\rm \tilde{t}_1}}
\newcommand{\mst}{m_{\rm \tilde{t}_1}}
\newcommand{\neu}{\tilde{\chi}^0}
\newcommand{\mneu}[1]{m_{\tilde{\chi}^0_{#1}}}

\newcommand{\gev}{\,\, \mathrm{GeV}}

\newcommand{\fbinv} {{\mathrm{fb}^{-1}}}
\newcommand{\Lum}   {{\cal{L}}}

\newcommand{\Ntracks}   {N_{\mathrm{tracks}}}
\newcommand{\Evis}      {E_{\mathrm{vis}}}
\newcommand{\pt}        {p_T}
\newcommand{\Njets}     {N_{\mathrm{jets}}}
\newcommand{\Pcharm}         {P_{\mathrm{c}}}
\newcommand{\Mjj}            {m_{\mathrm{jj}}}
\newcommand{\Mjjsq}          {m_{\mathrm{jj}}^2}
\newcommand{\gevsq}{\,\, \mathrm{GeV}^2}
\newcommand{\anc}{\rule{0mm}{0mm}}
\newcommand{\lesim}{\,\raisebox{-.1ex}{$_{\textstyle <}\atop^{\textstyle\sim}$}\,}

\begin{document}

\thispagestyle{empty}
\def\thefootnote{\fnsymbol{footnote}}       

\begin{center}
\mbox{ }

\end{center}

\vskip 3cm
\hspace*{-1cm}
\begin{picture}(0.001,0.001)(0,0)
\put(,0){
\begin{minipage}{\textwidth}
\begin{center}
\vskip 1.0cm

{\Huge\bf
Precision Measurements of the
}
\vspace{3mm}

{\Huge\bf
Stop Quark Mass at the ILC
}
\vskip 1.5cm
{\LARGE\bf A. Sopczak$^1$, A. Freitas$^2$, \\
\vspace{3mm}

          C. Milst\'ene$^3$, M.~Schmitt$^4$
\bigskip
\bigskip

\Large $^1$Lancaster University, UK; $^2$Zurich University, Switzerland; \\
       $^3$Fermilab, USA; $^4$Northwestern University, USA}

\vskip 2cm
{\Large \bf Abstract}

\end{center}
\end{minipage}
}
\end{picture}

\vskip 9.cm
\hspace*{-1cm}
\begin{picture}(0.001,0.001)(0,0)
\put(,0){
\begin{minipage}{\textwidth}
\Large
\renewcommand{\baselinestretch} {1.2}
Most supersymmetric models predict new particles within the 
reach of the next generation of colliders. For an understanding of 
the model structure and the mechanism(s) of electroweak symmetry 
breaking, it is important to know the masses 
of the new particles precisely. The measurement of the mass of the 
scalar partner of the top quark (stop) at an $\rm e^+e^-$ collider is 
studied. A relatively light stop is motivated by attempts to explain 
electroweak baryogenesis and can play an important role in dark 
matter annihilation. A method is presented which makes use of 
cross-section measurements near the pair-production threshold as well 
as at higher center-of-mass energies. It is shown that this method 
does not only increase the statistical precision, but also reduces 
the influence of systematic uncertainties, which can be important.
Numerical results are presented, based on a realistic event 
simulation, for two signal selection strategies: using conventional 
selection cuts, and using an Iterative Discriminant Analysis (IDA). 
While the analysis of stops is particularly challenging due to the 
possibility of stop hadronization and fragmentation, 
the general procedure could be applied to many precision mass measurements.
\renewcommand{\baselinestretch} {1.}

\normalsize
\vspace{2.5cm}
\begin{center}
{\sl \large
Presented at SUSY07, the 15th International Conference on Supersymmetry 
and the Unification of Fundamental Interactions, 
Karlsruhe, Germany, 2007, \\
to be published in the proceedings.
\vspace{-6cm}
}
\end{center}
\end{minipage}
}
\end{picture}
\vfill


\clearpage
\thispagestyle{empty}
\mbox{ }
\newpage
\setcounter{page}{0}
\pagestyle{fancyplain}

\title{Precision Measurements of the Stop Quark Mass at the ILC}
\author{Andr\'e Sopczak\inst{1}\thanks{\emph{Email: Andre.Sopczak@cern.ch} }
 \and
Ayres Freitas\inst{2}
\and
Caroline Milst\'ene\inst{3}
\and
Michael Schmitt\inst{4}
}                     
%
%
\institute{Lancaster University
\and Zurich University
\and Fermilab
\and Northwestern University
}
\date{}
\abstract{
Most supersymmetric models predict new particles within the 
reach of the next generation of colliders. For an understanding of 
the model structure and the
mechanism(s) of electroweak symmetry breaking, it is important to know the masses 
of the new particles precisely. The measurement of the mass of the 
scalar partner of the top quark (stop) at an $\rm e^+e^-$ collider is 
studied. A relatively light stop is motivated by attempts to explain 
electroweak baryogenesis and can play an important role in dark 
matter annihilation. A method is presented which makes use of 
cross-section measurements near the pair-production threshold as well 
as at higher center-of-mass energies. It is shown that this method 
does not only increase the statistical precision, but also reduces 
the influence of systematic uncertainties, which can be important.
Numerical results are presented, based on a realistic event 
simulation, for two signal selection strategies: using conventional 
selection cuts, and using an Iterative Discriminant Analysis (IDA). 
While the analysis of stops is particularly challenging due to the 
possibility of stop hadronization and fragmentation, the general procedure could be 
applied to many precision mass measurements.
} 
\maketitle

\section{Introduction}
\label{intro}

Supersymmetric particles are likely to be produced
and observed in high-energy proton-proton collisions at the LHC.  
However, it will be difficult to confirm their identity as
superpartners of the known Standard Model particles and to
measure their properties precisely.  For this, one needs 
experiments at a linear $\rm e^+e^-$ collider such as the proposed 
ILC at $\sqrt{s}=500$~GeV.

An experiment at the ILC will be able to make many precise 
measurements from which particle properties, and ultimately,
the outlines of a particle physics model may be inferred.
Due to the high statistical precision expected at the ILC, the optimization 
of the systematic errors is of particular importance.
We have studied one specific example, the extraction
of the mass of a scalar top quark from cross-section
measurements near threshold.  We have devised a method which
reduces most systematic uncertainties and leads to a potentially
very accurate measurement of the stop quark mass.  This method
is general and could be applied to other particles
that are pair-produced in an $\rm e^+e^-$ collider.

The method relies on the comparison of production rates at two different
center-of-mass energies, and knowledge of how the cross-section
varies as a function of $\sqrt{s}$ and the mass of the particle.

We have chosen the case of a light scalar top with
a mass not much higher than the mass of the lightest neutralino
since production of this particle was already extensively 
studied in an ILC context.
It was concluded that a conventional approach to the measurement of the
stop quark mass culminated in an uncertainty of about 1~GeV~\cite{carena}.
The new method improves substantially on this result.
The presented results are preliminary and being finalized~\cite{ayres}.

For this analysis, we have performed realistic simulations of the 
signal and backgrounds, and used two techniques to separate the 
signal from the background. The first technique is based on conventional 
sequential cuts, while the second employs an Iterative 
Discriminant Analysis (IDA). Furthermore, the hadronization followed by
fragmentation of the 
stop has been included and we have carefully studied the systematic 
uncertainties arising from this and other sources.

There are theoretical motivations for studying a light 
stop quark with a small mass difference.  Specifically, we
evoke a scenario within the Minimal Supersymmetric
extension of the Standard Model (MSSM) which is able to explain the
dark matter density of the universe as well as the baryon
asymmetry through the mechanism of electroweak baryogenesis~\cite{carena}.

A small mass difference between the stop and the lightest
neutralino can help to bring the dark matter relic density into the observed
region due to co-annihilation between the stop and the neutralino. For this
mechanism to be effective, the typical mass difference is rather small,
$\Delta m = \mst - \mneu{1} \lesim 30$ GeV.
The dominant decay mode of the stop is $\st \to c\,\neu_1$, resulting 
in a final state with two soft charm jets and missing energy.

Previous methods to determine the scalar top quark mass 
were discussed for the SPS-5 benchmark ($\mst=220.7$~GeV)~\cite{susy05}
and results are summarized in Table~\ref{tab:previous}.
For the cosmology motivated benchmark with $\mst=122.5$~GeV
and $\mneu{1} = 107.2$~GeV, 
an experimental precision of $\Delta \mst = \pm 1.0$~GeV was obtained~\cite{carena},
and about $\pm1.2$~GeV including theoretical uncertainties.
The following study investigates the same signal scenario and
it is based on the same background reactions and event preselection.

\begin{table}[htb]
\vspace*{-5mm}
\centering
\caption{Comparison of precision for scalar top mass determination
         for the SPS-5 benchmark ($\mst=220.7$~GeV).} 
\begin{tabular}{lcc}\hline
Method & $\Delta \mst$ (GeV) & luminosity \\
\hline
Polarization & 0.57 & $2 \times 500 \mathrm{\fb}^{-1}$ \\
Threshold scan & 1.2 & $300 \fb^{-1}$ \\
End point & 1.7 & $500 \fb^{-1}$\\
Minimum mass & 1.5 & $500 \fb^{-1}$ \\ \hline
\end{tabular}
\label{tab:previous}
\vspace{-0.9cm}
\end{table}

\section{Mass Determination Method}
\vspace{-1mm}
This method proposes to derive the stop mass from measurements at two center-of-mass
energies, one measuring the stop production cross-section near the threshold (th), and 
the other measuring it at a center-of-mass energy where the cross-section has approximately a 
peak (pk).
Using both measurements leads to a cancellation of systematic uncertainties in the mass
determination. A parameter $Y$ is defined as 
\begin{equation}
Y= \frac{N_{\rm th}-B_{\rm th}}{N_{\rm pk}-B_{\rm pk}}
=\frac{\sigma_{\rm th}}{\sigma_{\rm pk}}
\cdot\frac{\epsilon_{\rm th}}{\epsilon_{\rm pk}}
\cdot\frac{{\cal L}_{\rm th}}{{\cal L}_{\rm pk}},
\end{equation}
where $N$ is the total number of expected events after event selection and $B$ the number
of corresponding background events, $\sigma$ is the stop production cross-section,
$\epsilon$ the selection efficiency, and $\cal L$ the lumi\-no\-sity. The center-of-mass
energies 260 and 500~GeV have been chosen. Near the threshold, the production cross-section
is very sensitive to the stop mass.

In this study we assume that the ILC will operate primarily 
at $\sqrt{s}=500$~GeV with a total luminosity of ${\cal L} = 500$~fb$^{-1}$, 
and a small luminosity of 
${\cal L} = 50$~fb$^{-1}$ will be collected at $\sqrt{s}=260$~GeV. 
Table~\ref{tab:xsec} summarizes the expected production cross-sections.
The detector response was modeled with the SIMDET package including the LCFI vertex
detector concept.

\renewcommand{\arraystretch}{1.2}
\begin{table}[tp]
\centering
\tiny
\caption{Cross-sections for the stop signal and Standard Model background
processes for $\sqrt{s} = 260 \gev$ and $\sqrt{s} = 500 \gev$ and different
polarization
combinations. The signal is given for a right-chiral stop of $m_{\tilde{t}} =
122.5\gev$. 
Negative polarization values refer to
left-handed polarization and positive values to right-handed polarization.}
\label{tab:xsec}
\begin{tabular}{lrrrrrr}
\hline
Process &  \multicolumn{3}{c}{$\sigma$ (pb) at $\sqrt{s} = 260 \gev$} 
        &  \multicolumn{3}{c}{$\sigma$ (pb) at $\sqrt{s} = 500 \gev$} \\
\hline
$P(e^-) / P(e^+)$ \hspace*{-5mm} & 0/0 & -.8/+.6 & +.8/-.6
		  & 0/0 & -.8/+.6 & +.8/-.6 \\
\hline
$\tilde{t}_1 \tilde{t}_1^*$ & 0.032 & 0.017 & 0.077 & 0.118 & 0.072 & 0.276 \\
\hline
$W^+W^-$ & 16.9\phantom{0} & 48.6\phantom{0} & 1.77 
         & 8.6\phantom{0} & 24.5\phantom{0} & 0.77 \\
$ZZ$    & 1.12 & 2.28 & 0.99 & 0.49 & 1.02 & 0.44 \\
$W e\nu$ & 1.73 & 3.04 & 0.50 & 6.14 & 10.6\phantom{0} & 1.82 \\
$e e Z$  & 5.1\phantom{0} & 6.0\phantom{0} 
	 & 4.3\phantom{0} & 7.5\phantom{0} & 8.5\phantom{0} & 6.2\phantom{0} \\
$q \bar{q}$, $q \neq t$ & 49.5\phantom{0} & 92.7\phantom{0} & 53.1\phantom{0} 
			& 13.1\phantom{0} & 25.4\phantom{0} & 14.9\phantom{0} \\
$t \bar{t}$ & 0.0\phantom{0} & 0.0\phantom{0} & 0.0\phantom{0} & 0.55 & 1.13 &
0.50 \\
2-photon & 786\phantom{.00}&&&  \hspace*{-5mm} 936\phantom{.00}&& \\[-1ex]
$\pt > 5$ GeV\hspace*{-5mm} &&&&&& \\
\hline
\end{tabular}
\vspace{-0.2cm}
\end{table}
\renewcommand{\arraystretch}{1}

\begin{figure}
\begin{center}
\includegraphics[width=0.35\textwidth]{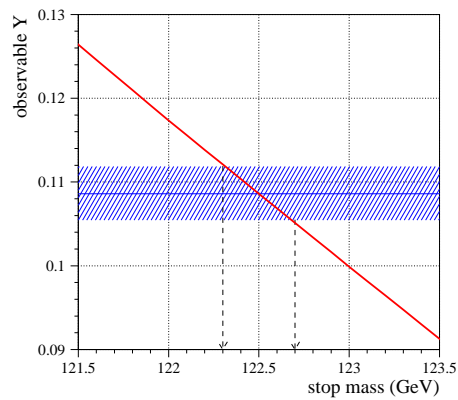}
\end{center}
\vspace*{-3mm}
\caption{Example of mass uncertainty derivation from the 
uncertainty of the observable $Y$.}
\label{fig:example}  
\end{figure}

The relation of the observable $Y$ and the stop mass is used to determine the stop mass
with precision. For example a variation of $Y$ by 3\% in a realistic scenario would
lead to an uncertainty of the stop mass $\Delta \mst = 0.2$~GeV as illustrated in 
Fig.~\ref{fig:example}. 

\vspace*{-2mm}
\section{Sequential-cut Analysis}
\vspace{-1mm}
In order to cancel the systematic uncertainties 
to a large extent with the described method,
the same sequential cuts are applied for the $\sqrt{s}=260$ and 500~GeV analyses. 
Details of the event selection are given in Table~\ref{tab:cuts} and the results are
given in Table~\ref{tab:nev}.

\renewcommand{\arraystretch}{1.2}
\begin{table}[hbp]
\centering
\tiny
\caption{Selection cuts for $\sqrt{s}=260$ and 500~GeV.
Also listed are the selection efficiencies optimized for right-chiral stop quarks.
}
\label{tab:cuts}
\begin{tabular}{lcc}
\hline
Variable &  $\sqrt{s} = 260 \gev$ 
         &  $\sqrt{s} = 500 \gev$ \\
\hline
number of charged tracks      
   &  $5 \le \Ntracks \le 25$ 
   &  $5 \le \Ntracks \le 20$ 
\\
visible energy $\Evis$
   &  $0.1 < \Evis/\sqrt{s} < 0.3$
   &  $0.1 < \Evis/\sqrt{s} < 0.3$
\\
event long. momentum
   &  $ |p_L / p_{\mathrm{tot}}| < 0.85$
   &  $ |p_L / p_{\mathrm{tot}}| < 0.85$
\\
event transv. momentum $\pt$
   &  $15 < \pt < 45$~GeV
   &  $22 < \pt < 50$~GeV
\\
thrust $T$
   &  $0.77 < T < 0.97$
   &  $0.55 < T < 0.90$
\\
Number of jets $\Njets$ 
   &  $\Njets \ge 2$ 
   &  $\Njets \ge 2$
\\
extra-jet veto
   &  $E_{\mathrm{jet}} < 25$~GeV
   &  $E_{\mathrm{jet}} < 25$~GeV
\\
charm tagging likelihood $\Pcharm$ 
   & $\Pcharm> 0.6$ 
   & $\Pcharm> 0.6$
\\
di-jet invariant mass $\Mjj$ 
   & $\Mjjsq < 5500 \gevsq$ or 
   & $\Mjjsq < 5500 \gevsq$ or 
\\

   & $\Mjjsq > 8000 \gevsq$
   & $\Mjjsq > 10000 \gevsq$
\\
\hline
signal efficiency & 0.340 & 0.212 \\
\hline
\end{tabular}
\end{table}
\renewcommand{\arraystretch}{1}

\renewcommand{\arraystretch}{1.2}
\begin{table}[tbhp]
\vspace*{-10mm}
\centering
\tiny
\caption{\label{tab:nev} 
Numbers of generated events, and expected events
for the sequential-cut analysis at $\sqrt{s} = 260$ and 500~GeV
for total luminosities of 50~fb$^{-1}$ and 200~fb$^{-1}$
with unpolarized and polarized beams.
}
\centering
\begin{tabular}{lrrrrrr}
\hline
 &  \multicolumn{3}{c}{$\sqrt{s} = 260\gev$} 
 &  \multicolumn{3}{c}{$\sqrt{s} = 500\gev$}\\
\hline
                  & generated & \multicolumn{2}{c}{$\Lum = 50~\fbinv$}
		  & generated & \multicolumn{2}{c}{$\Lum = 200~\fbinv$} \\
\hline
$P(e^-) / P(e^+)$ &   & 0/0 & {.8/-.6}
		  &   & 0/0 & {.8/-.6} \\
\hline
$\tilde{t}_1 \tilde{t}_1^*$ & 50,000 & 544 & 1309 
                            & 50,000 & 5170 & 12093  \\
\hline
$W^+W^-$ & 180,000 &   $38$ &   $4$ &   $210,000$ &   $16$ &  $2$  \\
$ZZ$     &  30,000 &    $8$ &   $7$ &    $30,000$ &    $36$ &   $32$ \\
$W e\nu$ & 210,000 &  $208$ &  $60$ &   $210,000$ & $7416$ & $2198$ \\
$e e Z$  & 210,000 &    $2$ &   $2$ &   $210,000$ &   $<7$ &  $<6$ \\ 
$q \bar{q}$, $q \neq t$ 
         & 350,000 &   $42$ &  $45$ &   $350,000$ &    $15$ &   $17$ \\
$t \bar{t}$ & ---  &    $0$ &   $0$ &   $180,000$ &    $7$ &   $7$ \\
2-photon & $1.6\times 10^6$ 
         & $53$ & $53$ & $8.5\times 10^6$ & $12$ & $12$  \\
\hline
\hspace*{-2mm} total background \hspace*{-2mm}& --- & $351$ & $171$ & --- & $7509$ & $2274$ \\
\hline
\end{tabular}
\vspace{-1.3cm}
\end{table}
\renewcommand{\arraystretch}{1}

\clearpage
\section{Iterative Discriminant Analysis (IDA)}
\begin{figure}[hp]
\vspace*{-12mm}
\begin{center}
\includegraphics[width=0.39\textwidth]{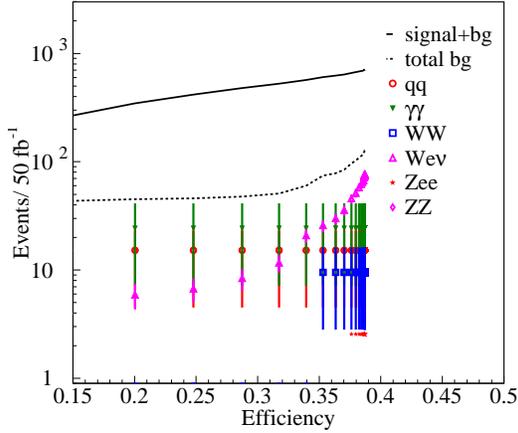}
\end{center}
\vspace*{-3mm}
\caption{IDA: Expected background events as a function of the signal efficiency for 
         ${\cal L} = 50$~fb$^{-1}$ at $\sqrt{s} = 260$ GeV.}
\label{fig:perf260}
\end{figure}

\begin{figure}[hp]
\vspace*{-16mm}
\begin{center}
\includegraphics[width=0.39\textwidth]{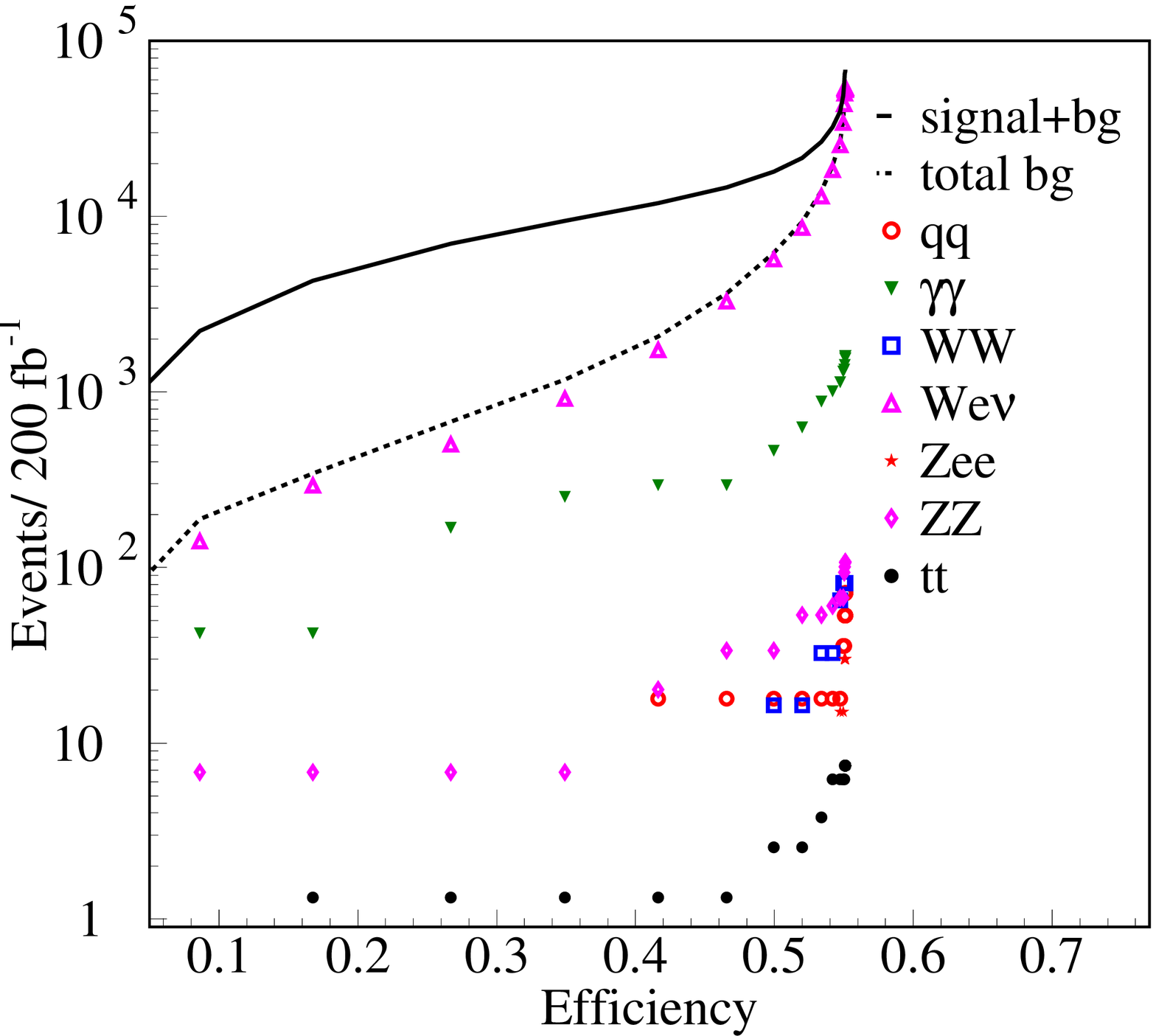}
\end{center}
\vspace*{-3mm}
\caption{IDA: Expected background events as a function of the signal efficiency for 
         ${\cal L} = 200$~fb$^{-1}$ at $\sqrt{s} = 500$ GeV.}
\label{fig:perf500}
\vspace*{-7mm}
\end{figure}

The Iterative Discriminant Analysis (IDA) is applied to increase
the discriminant power between signal and background
compared to the sequential-cut-based analysis, and thus reduce the 
statistical uncertainty in the stop mass measurement. Figures~\ref{fig:perf260}
and~\ref{fig:perf500} give the results of expected number of background events
as a function of the signal efficiency. The chosen working points have efficiencies
of 38.7\% and 41.6\% for the $\sqrt{s}=260$ and 500~GeV analyses, respectively.
Table~\ref{tab:ida2} lists the corresponding expected background events.

\renewcommand{\arraystretch}{1.2}%
\begin{table}[bhp]
\vspace*{-3mm}
\centering
\tiny
\caption{\label{tab:ida2}
Numbers of generated events, and expected events
for the IDA at $\sqrt{s} = 260$ and 500~GeV
for total luminosities of 50~fb$^{-1}$ and 200~fb$^{-1}$
with unpolarized and polarized beams.
}
\begin{tabular}{lrrrrrr}
\hline
 &  \multicolumn{3}{c}{$\sqrt{s} = 260\gev$} 
 &  \multicolumn{3}{c}{$\sqrt{s} = 500\gev$}\\
\hline
                  & generated & \multicolumn{2}{c}{$\Lum = 50~\fbinv$}
		  & generated & \multicolumn{2}{c}{$\Lum = 200~\fbinv$} \\
\hline
$P(e^-) / P(e^+)$\hspace*{-6mm} &   & 0/0 & {.8/-.6}
		  &   & 0/0 & {.8/-.6} \\
\hline
$\tilde{t}_1 \tilde{t}_1^*$ & 50,000 &   $619$  & $1489$
                            & 50,000 & $9815$  & $22958$  \\
\hline
$W^+W^-$ & 180,000 &   $11$ &   $1$ &   $210,000$ &   $<8$ &   $<1$ \\
$ZZ$     &  30,000 &   $<2$ &  $<2$ &    $30,000$ &    $20$ &   $18$ \\
$W e\nu$ & 210,000 &   $68$ &  $20$ &   $210,000$ &  $1719$ & $510$ \\
$e e Z$  & 210,000 &    $3$ &   $2$ &   $210,000$ &   $<7$ &  $<6$ \\ 
$q \bar{q}$, $q \neq t$ 
         & 350,000 &   $16$ &  $17$ &   $350,000$ &    $18$ &   $21$ \\
$t \bar{t}$ & ---  &    $0$ &   $0$ &   $180,000$ &     $1$ &    $1$ \\
2-photon & $1.6\times 10^6$ 
         & $27$ & $27$ & $8.5\times 10^6$ & $294$ & $294$  \\
\hline
\hspace*{-2mm} total background \hspace*{-6mm}& --- & $125$ & $67$ & --- & $2067$ & $851$ \\
\hline
\end{tabular}
\vspace*{-2mm}
\end{table}
\renewcommand{\arraystretch}{1}

\section{Systematic Uncertainties}
Both the sequential-cut-based analysis and the IDA method lead to a small 
statistical uncertainty resulting in \mbox{$\Delta \mst < 0.2$}~GeV and thus
systematic uncertainties are particularly important to evaluate.
Three classes of systematic uncertainties are distinguished:
\begin{itemize}
\item instrumental uncertainties related to the detector and accelerator:
      detector calibration (energy scale),
      track reconstruction efficiency,
      charm-quark tagging efficiency, and
      integrated luminosity.
\item Monte Carlo modeling uncertainty of the signal: charm and stop fragmentation effects.
      The Peterson fragmentation function was used with
      $\epsilon_{\rm c}=-0.031\pm 0.011$ (OPAL). For 
      $\epsilon_{\rm b}=-0.0041\pm 0.0004$ (OPAL) and 
      $\epsilon_{\rm b}=-0.0031\pm 0.0006$ (ALEPH) an average uncertainty of 15\% was 
      taken, and a factor 2 improvement at the ILC has been assumed, 
      leading to $\Delta\est=0.6\times 10^{-6}$ where 
      $\est=\epsilon_b(m_{\rm b}/m_{\rm \mst})^2$.
      Fragmentation effects and gluon radiation increase the number of jets significantly and
      the importance of c-quark tagging is stressed in order to resolve the combinatorics.
\item neutralino mass $108.2\pm 0.3$~GeV.
\item theoretical uncertainties on the signal and background. Some improvement compared
      to the current loop calculation techniques is assumed, and an even larger reduction of this uncertainty
      is anticipated before the start of the ILC operation.
\end{itemize}

Tables~\ref{tab:sys} and~\ref{tab:sysida} list the systematic uncertainties 
for the sequential-cut analysis and the IDA. The systematic uncertainty
using the IDA method from detector calibration (energy scale) is larger. 
This is because the sequential-cut analysis pays particular attention to cancellation 
of this uncertainty between the two \mbox{analyses} at the different center-of-mass energies. 

\renewcommand{\arraystretch}{1.2}
\begin{table}[hp]
\vspace*{-2mm}
\centering
\tiny
\caption{Sequential-cut analysis experimental systematic uncertainties
on the signal efficiency. The first column indicates the variable that is cut
on. The second column contains the expected systematic uncertainty for this variable 
based on experience from LEP.
The third column shows by how much the signal efficiency for
$\sqrt{s} = 260 \gev$ varies as a result of varying the cut value by this uncertainty. 
The fourth column shows the same for $\sqrt{s} = 500\gev$, and
the fifth column lists the resulting error on the observable $Y$.}
\label{tab:sys}
\centering
\begin{tabular}{lllll}
\hline
 & error on &  \multicolumn{2}{c}{rel. shift on signal eff.} & \\
variable & variable & $260 \gev$ & $ 500 \gev$ & error on $Y$ \\
\hline
energy scale    & $1\%$ &  $3.7\%$ &  $3.1\%$ & $0.6\%$ 
\\
$\Ntracks$      & $0.5\%$ &  \multicolumn{3}{c}{negligible}
\\
charm tagging   & $0.5\%$ &  \multicolumn{3}{c}{taken to be $0.5\%$}
\\
luminosity      & -- & $0.4\%$ & $0.2\%$ & $0.4\%$ 
\\
charm fragmentation & $0.011$ & $0.3\%$ & $0.8\%$ & $0.6\%$
\\
stop fragmentation & $0.6\times 10^{-6}$ & $0.6\%$ & $0.2\%$  & $0.7\%$ 
\\
neutralino mass & 0.3~GeV & 3.8\% & 3.0\% & 0.8\%
\\
background estimate & -- & 0.8\% & 0.1\% & 0.8\% 
\\
\hline
\end{tabular}
\end{table}
\renewcommand{\arraystretch}{1}

\renewcommand{\arraystretch}{1.2}
\begin{table}[bph]
\vspace*{-10mm}
\centering
\tiny
\caption{
IDA experimental systematic uncertainties.
\label{tab:sysida}
}
\centering
\begin{tabular}{lllll}
\hline
 & error on &  \multicolumn{2}{c}{rel. shift on signal eff.} & \\
variable & variable & $ 260 \gev$ & $ 500 \gev$ & error on $Y$ \\
\hline
energy scale    & $1\%$ &  $3.4\%$ &  $1.3\%$ & $2.3\%$ 
\\
$\Ntracks$      & $0.5\%$ &  \multicolumn{3}{c}{negligible}
\\
charm tagging   & $0.5\%$ &  \multicolumn{3}{c}{taken to be $0.5\%$}
\\
luminosity      & -- & $0.4\%$ & $0.2\%$ & $0.4\%$ 
\\
charm fragmentation & $0.011$ & $0.1\%$ & $0.6\%$ & $0.5\%$
\\
stop fragmentation & $0.6\times 10^{-6}$ & $0.1\%$ & $0.8\%$  & $0.7\%$ 
\\
neutralino mass & 0.3~GeV & 3.7\% & 1.6\% & 2.2\%
\\
background estimate & -- & 0.3\% & 0.2\% & 0.1\% 
\\
\hline
\end{tabular}
\vspace*{-8mm}
\end{table}
\renewcommand{\arraystretch}{1}
\clearpage

\section{Mass Determination}

The assessment of the achievable stop mass precision is based on the
statistical and systematic uncertainties on the observable $Y$ (eq. (1))
as summarized in Table~\ref{tab:sum}. 
The IDA method has a smaller statistical uncertainty, and also a smaller background 
uncertainty due to a smaller number of expected background events.
The expected stop mass uncertainty 
is inferred from the uncertainty on $Y$ as given in Table~\ref{tab:mstoperr}.

\renewcommand{\arraystretch}{1.2}
\begin{table}[hp]
\vspace*{-4mm}
\centering
\tiny
\caption{Summary of statistical and systematic uncertainties on
the observable~$Y$.
}
\label{tab:sum}
\centering
\begin{tabular}{lcc}
\hline
error source for $Y$ & sequential cuts & IDA method \\
\hline
statistical                           & 3.1\% &   2.7\% \\
detector effects                      & 0.9\% &   2.4\% \\
charm  fragmentation                  & 0.6\% &   0.5\% \\
stop fragmentation                    & 0.7\% &   0.7\% \\
neutralino mass                       & 0.8\% &   2.2\% \\
background estimate                   & 0.8\% &   0.1\% \\
\hline
sum of experimental systematics       & 1.7\% &   3.4\% \\
sum of experimental errors            & 3.5\% &   4.3\% \\
\hline
theory for signal cross-section       & 5.5\% &   5.5\% \\
\hline
total error $\Delta Y$                & 6.5\% &   7.0\% \\
\hline
\end{tabular}
\vspace*{-4mm}
\end{table}
\renewcommand{\arraystretch}{1}

\renewcommand{\arraystretch}{1.2}
\begin{table}[hp]
\vspace*{-4mm}
\centering
\tiny
\caption{Estimated measurement errors (in~GeV) on the stop quark mass.
\label{tab:mstoperr}}
\centering
\begin{tabular}{lcc}
\hline
 & \multicolumn{2}{c}{measurement error $\Delta\mst$ (GeV)} \\
error category & sequential cuts & IDA method \\
\hline
statistical                           & $0.19$   & $0.17$ \\
sum of experimental systematics       & $0.10$   & $0.21$ \\
beam spectrum and calibration         & $0.1\phantom{0}$    & $0.1\phantom{0}$  \\
sum of experimental errors            & $0.24$   & $0.28$ \\
sum of all exp. and th. errors
                                      & $0.42$   & $0.44$ \\
\hline
\end{tabular}
\vspace*{-5mm}
\end{table}
\renewcommand{\arraystretch}{1}

\section{Cold Dark Matter (CDM) Interpretation}
\begin{figure}[tb]
\begin{center}
\includegraphics[width=0.48\textwidth]{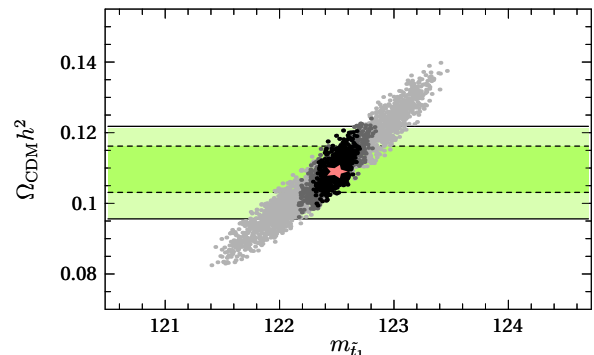}
\end{center}
\vspace*{-4mm}
\caption{
Computation of dark matter relic abundance $\Omega_{\rm CDM} h^2$
taking into account estimated experimental errors for stop, chargino, neutralino
sector measurements at the future ILC. The black dots correspond to
a scan over the 1$\sigma$ ($\Delta \chi^2 \leq 1$) region including the 
total expected experimental uncertainties (detector and simulation),
the grey-dotted region includes also the theory uncertainty,
and the light grey-dotted area are the previous results~\cite{carena}.
The red star indicates the best-fit point. 
The horizontal shaded bands show the
1$\sigma$ and 2$\sigma$ constraints on the relic
density measured by WMAP.}
\label{fig:stoppar}
\end{figure}

The chosen benchmark parameters are compatible with the mechanism of electroweak
baryogenesis~\cite{carena}.
They correspond to a value for the dark matter relic abundance
within the WMAP bounds, $\Omega_{\rm CDM} h^2 = 0.109$.
The relic dark matter density has been computed as in 
Ref.~\cite{carena}.
The assumed benchmark parameters changed slighty (larger slepton masses assumed) and
$\Omega_{\rm CDM} h^2$ changed from 0.1122~\cite{carena} to 0.109.
In the investigated scenario, the stop and lightest neutralino masses are 
$m_{\rm \tilde{t}_1} = 122.5~$GeV and $\mneu{1} = 107.2$~GeV, 
and the stop mixing angle is 
$\cos \theta_{\rm \tilde{\rm t}} = 0.0105$,
i.e. the light stop is almost completely right-chiral. 
The improvement compared to Ref.~\cite{carena} regarding the CDM precision 
determination is shown in Fig.~\ref{fig:stoppar} and summarized in Table.~\ref{tab:errors}.

\renewcommand{\arraystretch}{1.2}
\begin{table}[tb]
\caption{Estimated precision for the determination of stop mass and dark
matter relic density for different assumptions about the 
systematic errors.
}
\label{tab:errors}
\centering
\begin{tabular}{lcl}
\hline
 & $\Delta \mst$ (GeV) & $\Omega_{\rm CDM} h^2$ \\
\hline
exp. and th. errors   &
 0.42 & $0.109^{+0.015}_{-0.013}$ \\
stat. and exp. errors only && \\ 
\anc\hspace{1em} sequential-cut analysis &
 0.24 & $0.109^{+0.012}_{-0.010}$ \\
\anc\hspace{1em} IDA &
 0.28 & $0.109^{+0.012}_{-0.010}$ \\
\hline
\end{tabular}
\end{table}
\renewcommand{\arraystretch}{1}

\section{Conclusions}
Scalar top quarks could be studied with precision at a future International Linear Collider (ILC). 
The simulations for small stop-neutralino mass difference are motivated by cosmology.
The precision mass determination at the future ILC is possible with a method 
using two center-of-mass energies, e.g. $\sqrt{s}=260$ and 500~GeV.
This method can also be applied to other analyses to improve the mass resolution
in searches for new particles.
The precision of two independent analysis methods, one with a sequential-cuts and the
other with an Interative Discriminant Analysis (IDA) lead to very similar results.
The new proposed method increases the mass precision by about a factor of three due 
to the error cancellation using two center-of-mass energies with one near the production 
threshold.
Including experimental and theoretical uncertainties, the mass of a 122.5~GeV scalar top could be 
determined with a precision of 0.42~GeV. The interpretation of this benchmark scenario leads to a 
uncertainty on $\Omega_{\rm CDM} h^2$ of $-0.013$ and $+0.015$ which is about a factor two
better compared to previous results, and comparable to current cosmological (WMAP) 
measurement uncertainties.
With the new stop mass determination, the stop mass uncertainty is no longer the 
dominant uncertainty in the $\Omega_{\rm CDM} h^2$ calculation.


\end{document}